\documentclass[a4,twocolumn,pra,amsmath,amssymb,floatfix,groupaddress,superscriptaddress]{revtex4}
\usepackage{graphics}

\begin{document}

\title{Implementation of Spin Hamiltonians in Optical Lattices}

\author{J. J. \surname{Garc\'{\i}a-Ripoll}}
%\email{juan.ripoll@mpq.mpg.de}
\affiliation{Max-Planck-Institut f\"ur Quantenoptik, Hans-Kopfermann-Str. 1,
  Garching, D-85748, Germany.}
\author{M. A. \surname{Martin-Delgado}}
\affiliation{Max-Planck-Institut f\"ur Quantenoptik, Hans-Kopfermann-Str. 1,
  Garching, D-85748, Germany.}
\affiliation{Universidad Complutense de Madrid, Fac. de CC. F\'{\i}sicas,
  Ciudad Universitaria, Madrid, E-28040, Spain.}
\author{J. I. \surname{Cirac}}
\affiliation{Max-Planck-Institut f\"ur Quantenoptik, Hans-Kopfermann-Str. 1,
  Garching, D-85748, Germany.}

% 03.75.Kk Dynamic properties of condensates; collective and hydrodynamic
%         excitations, superfluid flow
% 03.75.Mn Multicomponent condensates; spinor condensates
% 75.10.Jm      Quantized spin models
% 32.80.Pj Optical cooling of atoms; trapping
% 03.75.Nt Other Bose-Einstein condensation phenomena
% 42.50.-p Quantum optics (for lasers, see 42.55.-f and 42.60.-v; see also
%         42.65.-k Nonlinear optics; 03.65.-w Quantum mechanics)
% 03.75.Lm Tunneling, Josephson effect, Bose-Einstein condensates in periodic
%         potentials, solitons, vortices and topological excitations


\date{\today}

\begin{abstract}
  We propose an optical lattice setup to investigate spin chains and ladders.
  Electric and magnetic fields allow us to vary at will the coupling
  constants, producing a variety of quantum phases including the Haldane
  phase, critical phases, quantum dimers etc.  Numerical simulations are
  presented showing how ground states can be prepared adiabatically.  We also
  propose ways to measure a number of observables, like energy gap, staggered
  magnetization, end-chain spins effects, spin correlations and the string
  order parameter.
\end{abstract}

\maketitle

In Condensed Matter Physics, there are strongly correlated systems of spins
and electrons whose study is extremely difficult both analytically and
numerically. These systems are of great practical interest since they may be
relevant in some important instances like high-Tc superconductivity, to quote
just one open problem of great relevance. These open problems have triggered a
great number of models formulated by means of quantum many-body Hamiltonians
like Heisenberg, t-J, Hubbard etc. Their quantum phase diagrams remain unknown
for generic values of coupling constants, electron concentration (doping) and
temperature, although a great knowledge can be obtained in particular
integrable models in one dimension.

Two very important examples of these systems are quantum spin chains and
ladders. Since the seminal work of Haldane \cite{Haldane} quantum spin chains
have been extensively studied as one of the simplest but most emblematic
quantum many-body systems.  According to Haldane, the one-dimensional
integer-spin Heisenberg antiferromagnets have a unique disordered ground state
with unbroken rotational symmetry and with a finite excitation gap in the
spectrum, while half-integer antiferromagnets are gapless and critical. This
quantum many-body phenomena is different from the usual source of gaps in
magnets, namely, single-ion anisotropy, which does not involve quantum
correlation effects.  Several theoretical developments \cite{Haldane, AKLT}
have helped to clarify the situation and there is now strong numerical
evidence in support of Haldane's claim.

\begin{figure}
  \centering
  \resizebox{0.45\linewidth}{!}{\includegraphics{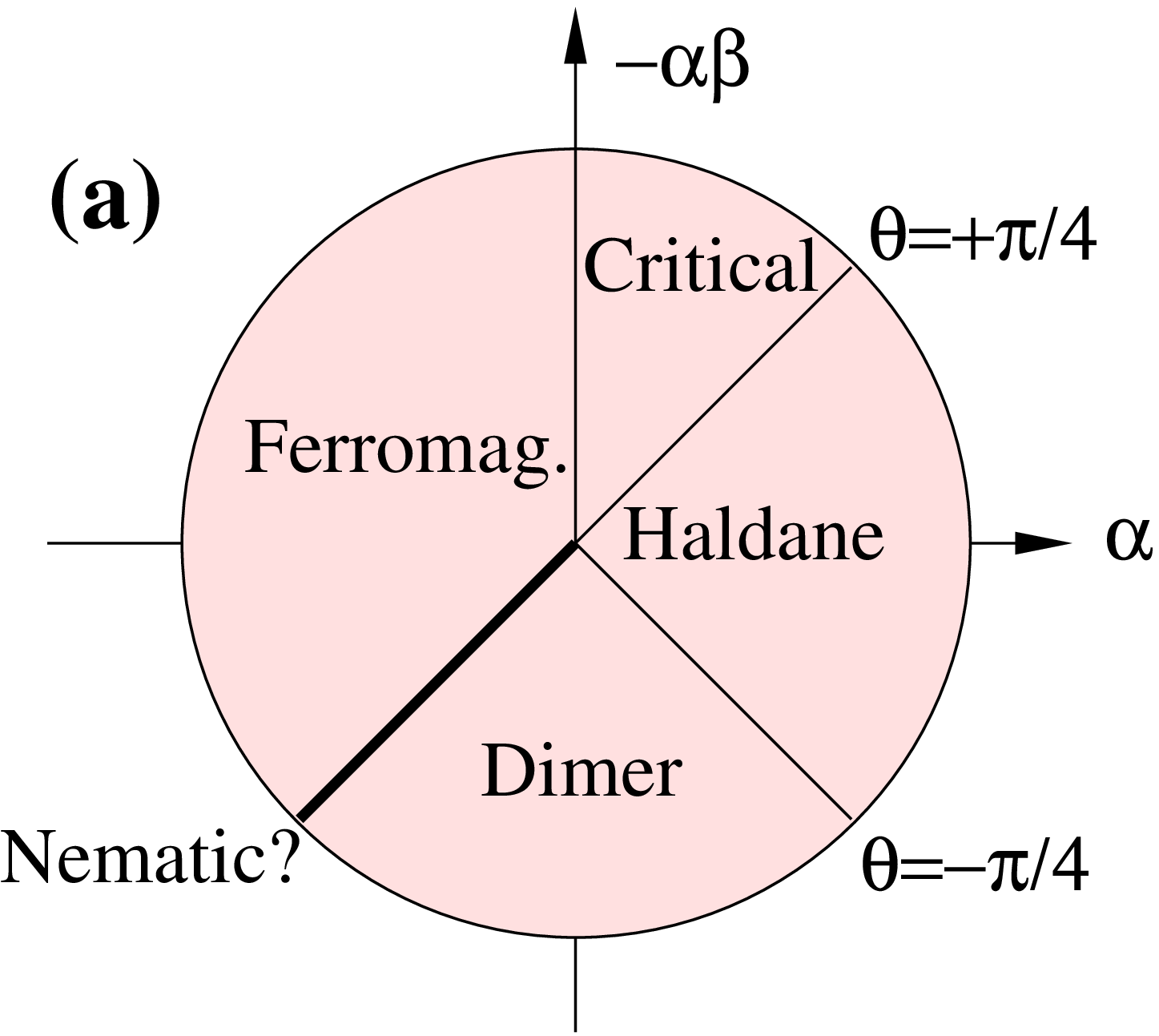}}
  \resizebox{0.47\linewidth}{!}{\includegraphics{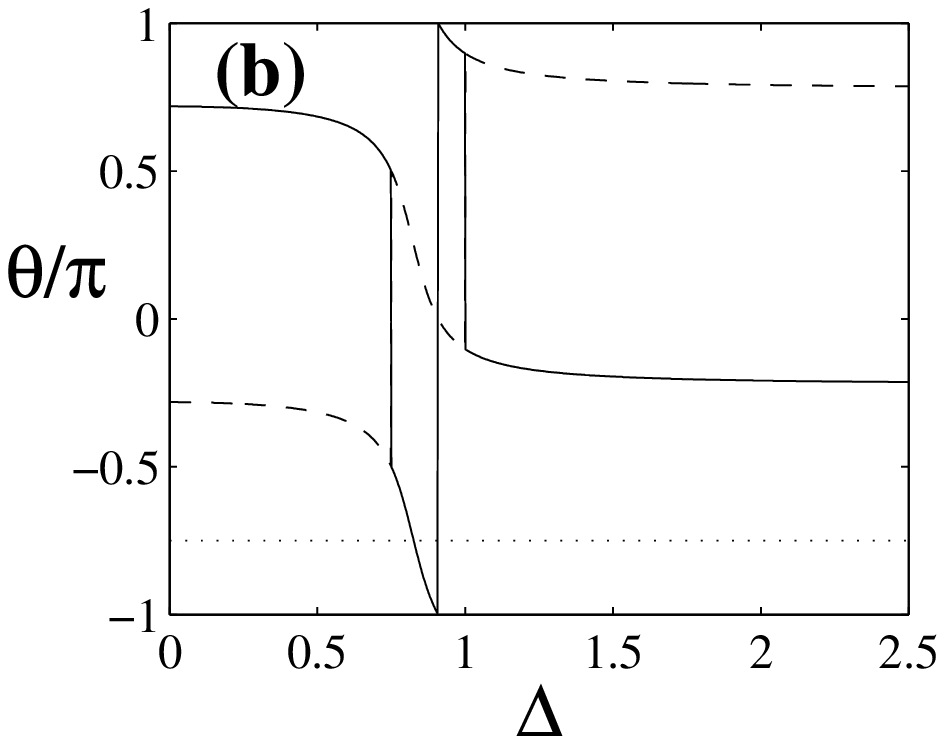}}
  \caption{
    (a) Different phases of the ground state of Hamiltonian (\ref{H-spin}),
    according to the ``angle'' $\theta$ defined in the text. (b) Type of
    Hamiltonian (\ref{parameters}) as a function of the gradient of the
    electric field, $\Delta$, for $U_0=0.75 U_2$. The solid line is obtained
    naturally, the dashed line is for the dual model obtained when working on
    the upper part of the spectrum, and the dotted line marks the location of
    the possible nematic phase.}
  \label{fig-delta}
\end{figure}

Here we shall address the issue of implementing spin chains and ladders in an
optical lattice. For concreteness, we focus on spin chains first.  For integer
spin, $s=1$, there is a quantum Hamiltonian that contains all the relevant
information pertaining the Haldane phase and exhibiting a rich phase
structure. This is the so--called Quadratic-Biquadratic Hamiltonian (QBH)
given by
\begin{equation}
H_{\rm QB} = \alpha \sum_{i=1}^{N-1}\left[
  \vec{S}_{i}\cdot \vec{S}_{i+1} -
  \beta (\vec{S}_{i}\cdot \vec{S}_{i+1})^2 \right] + \sum_{i=1}^{N} \vec{B}_i\vec{S}_i.
\label{H-spin}
\end{equation}
Here, $\vec{S}_{i}$ are spins $s=1$ at lattice site $i$, $\beta$ is a relative
coupling constant that parametrizes a family of local Hamiltonians, and the
sign of $\alpha$ determines the ferro or antiferromagnetic regimes. The
properties of the ground state without magnetic field, $\vec{B}=0$, are
entirely determined by an angle, $\theta$, such that $\alpha =
|a|\cos(\theta)$ and $\alpha \beta = -|a|\sin(\theta)$ [See
Fig.~\ref{fig-delta}(a)]. For $\theta \in [-\frac{\pi}{4},\frac{\pi}{4}]$, the
ground state belongs to the Haldane phase, with $\theta =0$ being the
Heisenberg point and $\theta =\arctan(\frac{1}{3})$ the AKLT point
\cite{AKLT}, which is of particular importance because it can be described
with an exact valence--bond wavefunction. There are two critical points on
which the standard correlation length $\xi_{\rm C}$ diverges: $\theta
=\frac{\pi}{4}$ is the Uimin-Lai-Sutherland (ULS) critical point \cite{ULS},
where a phase transition occurs into a gapless phase for $\theta \in
[\frac{\pi}{4},\frac{\pi}{2}]$; and $\theta =-\frac{\pi}{4}$ corresponds to
the Takhtajan-Babujian (TB) critical point \cite{TB} displaying a second order
phase transition into a dimerized phase for $\theta \in
[\frac{-3\pi}{4},-\frac{\pi}{4}]$, thus gapped.  It has been conjectured that
a quantum nondimerized nematic phase also exists \cite{chubukov} in the
ferromagnetic region.

Likewise, ladders of spin $s=\frac{1}{2}$ \cite{ladders} exhibit a rich
quantum phase structure depending on their number of legs and couplings: for
even legs, ladders are gapped and can be in a Haldane phase similar to the
integer spin chains, whereas for odd legs they are gapless.  Even-legged
ladders can also be in quantum dimer phases located on the rungs, provided the
vertical couplings are strong enough.  Theoretically, spin ladders are
regarded as a route to approach the more complicate physics of two-dimensional
quantum spin systems, as we increase the number of legs. For instance, a
two-leg ladder is gapped and upon hole doping can serve as a toy model for
studying superconducting correlations.

Experimental attempts to implement the QB Hamiltonian in real materials date
back to the sixties. The first experimental evidence of a biquadratic term was
found in Mn-doped MgO \cite{sixties} and for $S=5/2$ $\rm Mn^{++}$ ions in
antiferromagnetic chains of MnO \cite{sixties}. The measured coupling constant
was $\beta=-0.05$, too small and with a fixed sign. Thus, only the pure
Heisenberg AF model and its immediate surroundings with a very small $\beta$
are of feasible practical implementation, and the AKLT point is rather far to
be accessible. Experimentally, ladders can be realized by selective
stoichiometric composition of cuprate planes in superconductors \cite{GRS}.
However, residual non-vanishing inter-ladder couplings on the planes introduce
disturbances which are difficult to control.

Finding an experimental setup for checking the validity and observing the
several phases of the QB Hamiltonian and of spin ladders is considered a very
important challenge in the field. In this paper we propose to solve this
problem using cold atoms confined in an optical lattice \cite{Bloch}.  As
shown before, a Mott phase \cite{bh} of cold atoms in a lattice can be
described using ferromagnetic spin $s=\frac 1 2$ \cite{Duan,njp} or $s=1$
\cite{Yip03,Lukin} Hamiltonians. Here we describe how to access a wider family
of models, including Haldane phases of antiferromagnetic $s=1$ chains and
$s=\frac{1}{2}$ ladders.  We also design a technique to prepare adiabatically
the atoms in the ground state, an important task since these spins cannot be
cooled.  Finally we study how to detect the different spin phases, and to
directly observe correlation and excitation properties.  This is in sharp
contrast with standard experiments in condensed matter where one neither has a
controllable implementation of the QB Hamiltonian (\ref{H-spin}) nor a direct
way to perform measurements that so far have been regarded as mere theoretical
tools.

Let us first consider how to engineer Hamiltonian (\ref{H-spin}) using spin
$s=1$ bosons in an array of one-dimensional optical lattices.  For a strong
confinement and low densities, the effective Hamiltonian is the Bose-Hubbard
model \cite{bh}
\begin{eqnarray}
  H &=& -J \sum_{\langle{j,l}\rangle,\alpha} (a^\dagger_{j\alpha} a_{l\alpha}+
  a^\dagger_{l\alpha} a_{j\alpha}) +  \sum_{j,\alpha} (E_j + B_{j\alpha}) a^\dagger_{j\alpha}a_{j\alpha}\nonumber\\
  &+&\sum_{S=0,2} \frac{U_S}{2} \sum_{j,\alpha,\beta,\gamma,\delta}
  (\Psi^{(S)}_{\sigma,\gamma\delta}a_{j\gamma}a_{j\delta})^\dagger
  (\Psi^{(S)}_{\sigma,\alpha\beta}a_{j\alpha}a_{j\beta}).
  \label{BH}
\end{eqnarray}
First of all, while the indices $j$ and $l$ run over the lattice sites, the
Greek letters label the projection along the Z axis of either the spin of an
atom ($\alpha, \beta, \gamma, \delta=-1,0,+1$), or of a pair of them
($\sigma=-2,-1,0,1,2$). Then the first term in the Hamiltonian is the
single-particle hopping term and $J$ is the tunneling amplitude to a
neighboring site. The second term models the interaction between bosons in a
single site: two bosons can only interact if their total spin is either $0$ or
$2$, because the state $S=1$ is antisymmetric.  Furthermore, the interaction
may be different for each value of the total spin. Both statements are
summarized in the presence of spin-dependent interaction constants, $U_S$, and
in the tensors $\Psi^{(S)}_{\sigma,\gamma\delta} = \langle{S,\sigma|s,\gamma;
  s, \beta}\rangle$, which are the Clebsch-Gordan coefficients between the
states $|s=1,\gamma\rangle\otimes|s=1,\beta\rangle$ and
$|S=0,2;\sigma\rangle$.  Finally, we have included effective electric and
magnetic fields, $E_j$ and $B_{j\alpha}$, that can be engineered using Stark
shifts and spatially dependent magnetic fields, as in current experiments
\cite{Bloch}.

\begin{figure}[t]
  \centering
  \resizebox{\linewidth}{!}{\includegraphics{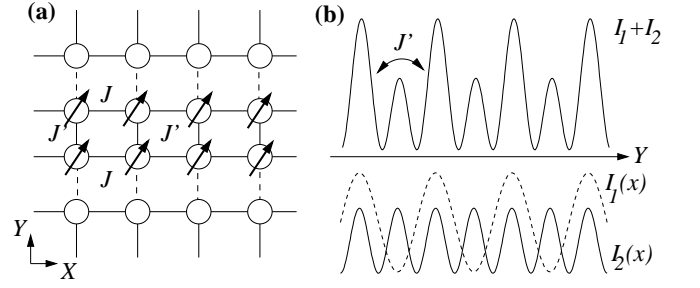}}
  \caption{
    (a) A ladder is the combination of two spin chains that interact with each
    other. Interactions along a chain and between legs can be different. (b)
    We can build a ladder combining a 3D lattice, $I_2(x)$, with an additional
    1D optical lattice, $I_1(x)$, that has twice the period of the first one.
    This induces a tunable hopping $J'$, different from the longitudinal one,
    $J$, and suppresses hopping between neighboring ladders.}
  \label{fig-ladder}
\end{figure}

We will assume that the lattice has been loaded with one atom per site
\cite{filtering}, and that the tunneling has been strongly suppressed, $J \ll
U_S$. With a perturbative calculation around states with unit occupation
\cite{njp,Yip03,Lukin} we obtain the QB Hamiltonian (\ref{H-spin}) with
constants
\begin{equation}
  \alpha = \tfrac{1}{2}C_2,\; \alpha\beta = -\tfrac{1}{6}(2C_0+C_2),\;
  C_S=\frac{J^2U_S}{\Delta^2-U_S^2}
  \label{parameters}
\end{equation}
This result is valid only if the gradient of the magnetic field is small,
$|B_{j+1}-B_j| \ll |U_S|$, and the gradient of the electric field is constant,
$\Delta = E_{j+1}-E_j$, and not resonant with the interaction,
$|nU_S\pm\Delta| \gg J,\; \forall n \in\mathbb{Z}$.

\begin{figure}
  \centering
  \resizebox{\linewidth}{!}{\includegraphics{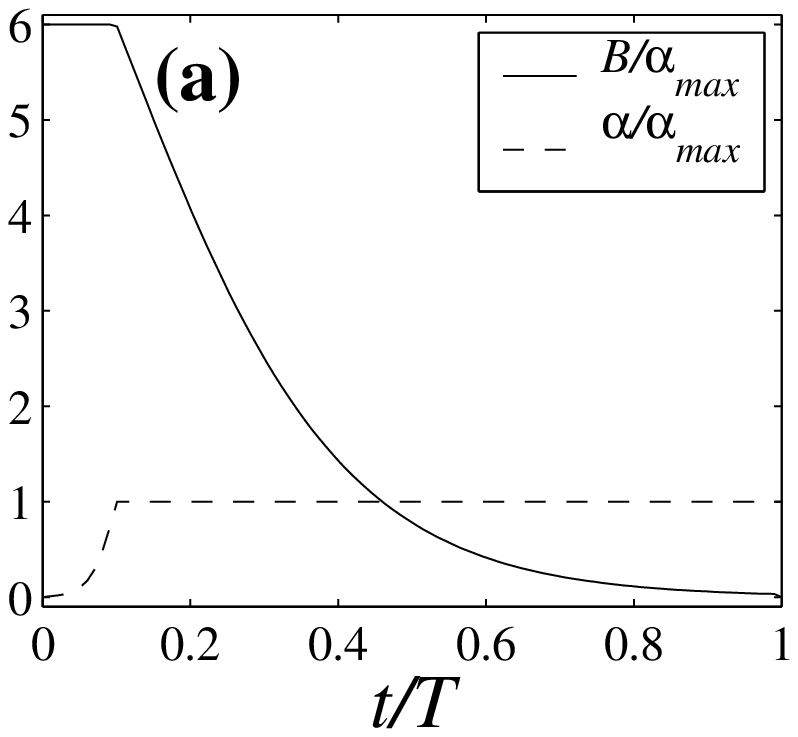}
    \includegraphics{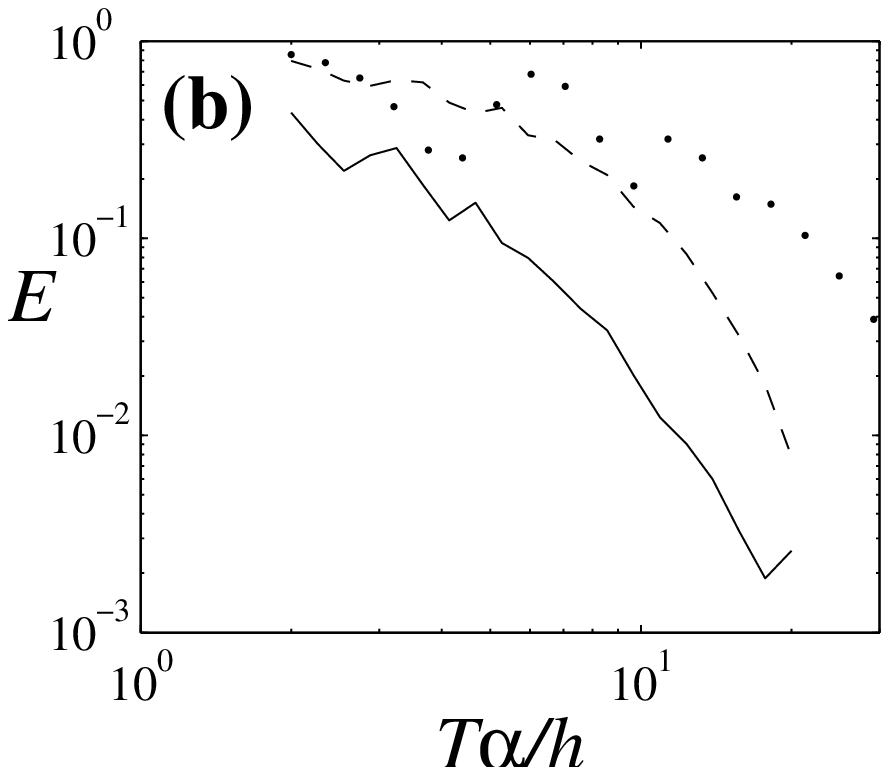}}
  \caption{(a) Procedure for the adiabatic construction of the ground
    state of the $s=1$ antiferromagnet with open boundary conditions.  (b)
    Infidelity of the final state for the AKLT (dashed), Heisenberg (solid)
    and $s=\frac 1 2$ spin ladder (dots), as a function of the duration of the
    process, for a setup with 9 spins (chains) or rungs (ladders).}
  \label{fig-fidelity}
\end{figure}

In the absence of electric or magnetic fields the model reduces to that of
\cite{Yip03} and \cite{Lukin}, and we are restricted to a fixed value of
$\theta$, typically in the ferromagnetic sector. However, with our tools it is
possible to explore many other phases and achieve almost all values of
$\theta$ [Fig.~\ref{fig-fidelity}]. The idea is to change the gradient of the
electric field and use a duality $H_{AF} = - H_{F}$ between ferro and
antiferromagnetic models: The highest energy state of a ferromagnetic model
($\alpha < 0$) is the same and exhibits the same dynamics as the ground state
of the dual model ($-\alpha, -\beta$), since $ \left(i\partial_t -
  H_F\right)\psi(t) = 0 \Leftrightarrow \left(i\partial_t -
  H_{AF}\right)\psi^\star(t)=0$. This equivalence is possible in current
experiments, because dissipation is negligible and decoherence affects equally
both ends of the spectrum.

A similar procedure is used for implementing ladders \cite{ladders} of spin
$s=\frac 1 2$. A ladder is nothing but the combination of two spin chains
(legs) that interact with each other. To build them we need to set up a 3D
lattice that confines the atoms on planar square lattices (hopping has been
suppressed along the $Z$ direction), and superimpose along the $Y$ direction a
second 1D lattice with twice the period [Fig.~\ref{fig-ladder}(b)]. Adjusting
the intensities of different lattices we can modify the tunneling along the
leg of a ladder and between neighboring legs, and completely suppress
tunneling between ladders.  With the help of electric and magnetic fields
\cite{njp}, and the duality between ferro and antiferromagnetic models, we
achieve once more a full tunability of the Hamiltonian.

Let us now study how to prepare ground states adiabatically. We will focus on
the Haldane phase of the $s=1$ lattice and on the antiferromagnetic $s=\frac 1
2$ ladders. Since in both cases we seek an antiferromagnetic state, we can
begin with a configuration of antiparallel spins, an effective staggered
magnetic field, $B_j = (-1)^j|B(t)|$, and no hopping. We then progressively
decrease the magnetic field and increase the interaction, $\alpha$ [See Fig.
\ref{fig-fidelity}(a)].  This procedure constrains us to a subspace of fixed
magnetization, $\langle{\sum_jS^z_j}\rangle\simeq 0$, and also ensures that
the minimum energy gap between the ground state and the first excitations
remains independent of the number of spins. Thus, the speed of the adiabatic
process can be the same for all lattice sizes, an important point in a setup
with defects.

We have studied numerically the fidelity of the adiabatic process for the AKLT
point, for $\beta=0$, and for a $s= \frac 1 2$ ladder, using different speeds
and sizes. The fidelity is the projection of the final state onto the
(degenerate) ground states (in the case of $s=1$, it is one singlet and three
triplet states).  The results are shown in Fig.  \ref{fig-fidelity}(b).  Already
for a duration of $10 \times h/\alpha_{max}$ the Haldane phase is built with
$99\%$ fidelity.  Assuming current optical lattices, with $U/h\simeq 3$KHz and
$J=0.3U$, this implies a time of roughly $37$ ms. Improvements on these values
are expected with the implementation of optically induced Feschbach resonances
\cite{feschbach}.

\begin{figure}[t]
  \centering
  \resizebox{0.98\linewidth}{!}{\includegraphics{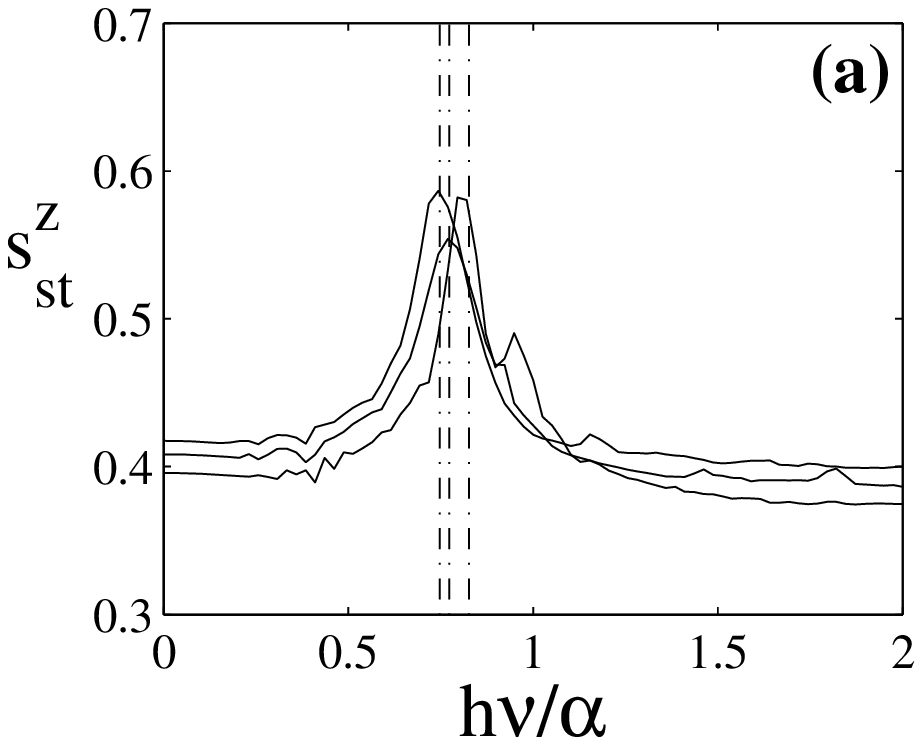}
    \includegraphics{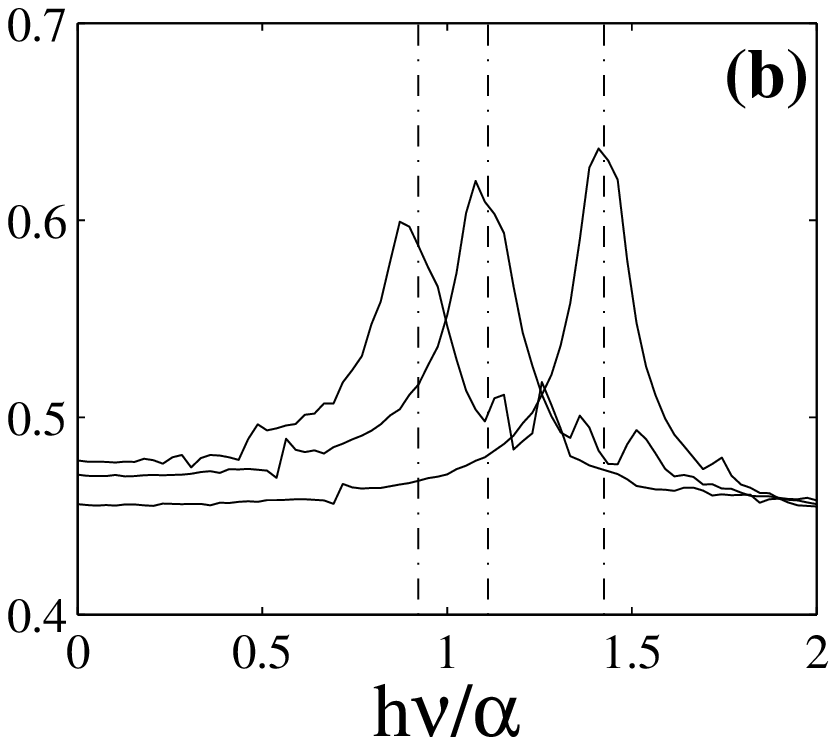}
    }
  \caption{
    Maximum staggered magnetization per spin acquired by the ground state of
    the (a) AKLT and (b) Heisenberg models, under an oscillating magnetic
    field, $B_j = (-1)^j 0.025 \alpha \cos(2\pi \nu t)$, for $5$, $7$ and $9$
    spins (right to left). Vertical lines mark the finite excitation gap for
    the given size.}
  \label{fig-reson}
\end{figure}

Regarding the staggered magnetic field, for $^{87}$Rb in the $F=1$ hyperfine
state it can be produced using a weaker optical lattice, aligned with the
atomic chains and made of a pair of counterpropagating laser beams in a lin
$\perp$ lin configuration. If the lasers are far off-resonance from the
transition $^2\mathrm{S}\to{}^2\mathrm{P}$, we will obtain a state-dependent
potential $V_\perp(x) \propto 2 + (\sin(kx)^2-\cos(kx)^2)S_z$, where the
$\sin(kx)$ and $\cos(kx)$ come from the Stark shifts induced by the $\sigma_+$
and $\sigma_-$ polarizations on the atomic states.  Choosing the orientation
of the counterpropagating beams so that $V_\perp(x)$ has twice the periodicity
of the confining lattice we get our staggered magnetic field. A similar setup
can be designed for $s=\frac 1 2$ particles.

Once we have constructed the ground state, we would like to study its
properties. We will describe a number of possible experiments, sorted by
increasing difficulty. For illustrative purposes, we consider the Haldane
phase of a $s=1$ spin chain, but we want to emphasize that the same techniques
can be applied to other phases, spin models and even ladders.  Preliminary
evidences of the Haldane phase can be obtained by studying global properties,
such as the staggered magnetization, $\vec{S}_{st}=\sum_j(-1)^j\vec{S}_j$,
which is zero in the dimer phase and nonzero in the Haldane phase. To measure
$S_{st}^{x,y}$, we apply a $\pi/2$ rotation around the $Z$ axis using the
staggered magnetic field, and then measure $\langle \sum_i S_i^{x,y}\rangle$.
The remaining component $S_{st}^z$ can be obtained by rotating all spins an
angle $\pi/2$ around the $X$ or $Y$ axes, and then measuring $S_{st}^y$ or
$S_{st}^x$.

We can also study the energy gap between the ground state and its excitations,
using an oscillating magnetic field, $B_j(t) = (-1)^j B \sin(\omega t)$. From
linear response theory we know that for small intensities, $|B|\ll\alpha$,
there is a strong resonance at the gap, $\hbar\omega= E_{gap}$, which
manifests itself on the growth of the staggered magnetization, $S^z_{st}$.
This result has been confirmed by numerical simulations of small lattices, as
shown in Fig.  \ref{fig-reson}.

Another interesting feature of the Haldane phase are fractionalization
effects. To understand this, one should visualize each atom with spin $s=1$ as
being composed of two $s=\frac 1 2$ bosons in a symmetric state. The ground
state of Eq. (\ref{H-spin}) can then be built ---either approximately, if
$\beta\neq1/3$, or exactly, for the AKLT---, by antisymmetrizing pairs of
virtual spins from neighboring sites \cite{AKLT}. This leaves us with two free
effective $s=\frac 1 2$ spins at the ends of a chain, which manifest
themselves physically.  First, the four almost degenerate ground states are
determined by the values of the free virtual spins, which we will denote as
$\left|i_1,i_{2N}\right\rangle$.  Thus, if the state of the system is
$\left|\psi\right\rangle=\sum_{ij}c_{ij}\left|i,j\right\rangle$, the
probability of measuring the left- and rightmost real spins in states
$I,J=\pm1$ is approximately $4|c_{I/2,J/2}|^2/9$.  And second, the virtual
spins almost do not interact and can be manipulated independently with weak
magnetic fields that have different values on the borders of a chain.  For
instance, if we prepare the ground state using our method, apply a global
rotation of angle $\theta=\frac \pi 2$ around the $Y$ axis, and then switch on
the staggered magnetic field, we will measure periodic oscillations in the
value of $ S^x_{st}$, because the virtual spins on even and odd sites rotate
with opposite senses.

We also have developed a procedure to measure spin correlation functions in
optical lattices, an essential tool for experiments with spin lattices. Our
proposal only assumes that we can trap atoms in separate lattices and empty
lattice sites with double occupation \footnote{This may be done, for instance,
  photoassociating atoms from a single site to form a molecule.}.  First of
all, we notice that a spin correlation may be written as a density
correlation, $\langle S^z_j S^z_{k}\rangle = \langle n^{+1}_j n^{+1}_k +
n^{-1}_j n^{-1}_k - n^{+1}_j n^{-1}_k - n^{-1}_j n^{+1}_k\rangle$, where
$n^\alpha_k$ is the number of atoms in hyperfine state $\alpha$ on the $k$-th
site. A correlation such as $\sum_j\langle n^1_j n^1_{j+\Delta}\rangle$ can be
measured by moving the atoms of species $+1$ just $\Delta$ sites
\cite{qc,bloch-qc}, emptying all doubly occupated sites and counting the
number of atoms left in state $+1$, on the sites $j$ and $k$.  If we rather
count the atoms in state $0$, then we will obtain $\langle n^{-1}_j
n^{+1}_k\rangle$.  With this method, and some global rotation of the spins, it
is possible to obtain all correlators, $\langle \vec{S}_j
\vec{S}_{j+\Delta}\rangle$. If we cannot address individual atoms, using the
same procedure and measuring total populations, $\sum_k n^\alpha_k$, we will
obtain averaged values, $\sum_k\langle \vec{S}_j \vec{S}_{j+\Delta}\rangle$.
Both quantities are interesting to discriminate between the different spin
phases.

On a much higher level of difficulty, but with pretty much the same tools as
for measuring correlations we can obtain the \textit{string order parameters}
\cite{string-order}. One can show that it is equivalent to a correlation
function measured on a transformed state,
\begin{subequations}
\begin{eqnarray}
  {\cal S}_{km} &:=&
  \left\langle S^z_k e^{i \pi\sum_{j=k+1}^{m-1} S^z_j} S^z_m \right\rangle_\psi
  = \left\langle S^z_k S^z_m \right\rangle_{U_{m-k} \psi},\\
  U_\Delta &=&  \exp
  \left[i \pi \sum_{k=1}^{N}
    \sum_{j=k+1}^{k+\Delta} \left(1-{S^z_k}^2\right) S^x_j\right]
\end{eqnarray}
\end{subequations}
The unitary operation $U_\Delta$ can be performed as follows. First we
perform half a swap between the $+1$ and $-1$ states, $U_1 = \exp\left[i
  \pi/2\sum_k(\left|+1\right\rangle\left\langle-1\right|_k+
  \left|-1\right\rangle\left\langle+1\right|_k)\right]$, with a $\pi/2$
Raman pulse that connects these states. Next we split the three species
into three optical lattices. The atoms in state $0$ will move $\Delta-1$
sites to the right, and on each movement a controlled collision with atoms
in state $+1$ will take place. Adjusting the duration of this collision so
that it produces a phase of $\pi$, we obtain the transformation $U_2 =
\exp\left[i \sum_k \sum_{j=1}^{\Delta-1}
  (\left|0\right\rangle\left\langle0\right|_k
  \left|+1\right\rangle\left\langle+1\right|_{k+j})\right]$. We restore all
atoms back to their positions, and repeat the operation $U_1$, concluding
the total transformation $U_\Delta = U_1 U_2 U_1$.  Finally, we perform all
required steps to measure either $\langle S^z_k S^z_{k+\Delta}\rangle$ or
$\sum_k\langle S^z_k S^z_{k+\Delta}\rangle$.

Typical experiments have defects and thus host chains with different number
of spins. However, except for the string-order parameter, the measurements
that we propose are extremely robust, and in general they produce a signal
that is a nonzero average of the possible outcomes for chains of different
lengths.

Summing up, in this work we have shown how to implement spin $s=1$ chains and
$s=\frac 1 2$ ladders with cold atoms in an optical lattice. Such experiments
will allow us to construct never observed phases and probably dilucidate the
existence of the nematic phase. Finally, we have developed a very general set
of tools to characterize these spin phases, which are themselves of interest
for future experiments with optical lattices.

This work has been supported by the EU projects TOPQUIP and CONQUEST,
\textit{Kompetenz\-netz\-werk Quanten\-informations\-verarbeitung der
  Bayerischen Staatsregierung}, SFB631, and DGES under Contract
BFM2000-1320-C02-01.

\end{document}